\documentclass[twocolumn,aps,showpacs,amsmath,amssymb]{revtex4}

\usepackage{graphicx}
\usepackage{dcolumn}
\usepackage{bm}
\usepackage[T1]{fontenc}
\usepackage{ae,aecompl}

\begin{document}

\preprint{APS/123-QED}

\title{Self diffusion of reversibly aggregating spheres}
\author{Sujin Babu}
\author{Jean Christophe Gimel}
\email{Jean-Christophe.Gimel@univ-lemans.fr}
\author{Taco Nicolai}%
\affiliation{Polym\`eres Collo\"{\i}des Interfaces, CNRS UMR6120,
Universit\'e du Maine, F-72085 Le Mans cedex 9, France
}%

\date{\today}
\begin{abstract}
Reversible diffusion limited cluster aggregation of hard spheres
with rigid bonds was simulated and the self diffusion coefficient
was determined for equilibrated systems. The effect of increasing
attraction strength was determined for systems at different volume
fractions and different interaction ranges. It was found that the
slowing down of the diffusion coefficient due to crowding is
decoupled from that due to cluster formation. The diffusion
coefficient could be calculated from the cluster size distribution
and became zero only at infinite attraction strength when
permanent gels are formed. It is concluded that so-called
attractive glasses are not formed at finite interaction strength.
\end{abstract}

\maketitle
\section{Introduction}
Reversible aggregation of small particles in solution is a common
phenomenon. It leads to different equilibrium states depending on
the volume fraction ($\phi$) of the particles and the strength of the
interaction energy ($u$). Weak attraction results in the formation of transient
aggregates at low $\phi$  and a transient percolating network at high $\phi$,
while strong attraction may drive phase separation into a high and
a low density liquid 
\cite{433,867,823,869,652,488,913,489,288,896,791,834,878,833,790,876,898,926,907,899,938,891,927,931,933,912,883,910,955,831,832}. 
The strength of the interaction and thus the
equilibrium properties are determined by the ratio of the bond
formation ($\alpha$) and the bond breaking ($\beta$) probability \cite{943,942,545,413,270,269,267,411,415,478,754,725,753,901}, while the
kinetics of such systems depend on the absolute values of $\alpha$ and $\beta$.
Two limiting cases may be distinguished: diffusion limited cluster
aggregation (DLCA) for which a bond is formed at each collision ($\alpha
=1$) and reaction limited cluster aggregation (RLCA) for which the
probability to form bonds goes to zero ($\alpha\rightarrow 0$). 

The average long time self diffusion coefficient ($D_l$) of
non-interacting hard spheres decreases with increasing volume
fraction \cite{877,872,1009,1010,993} and the system forms a glass above a volume
fraction of about 0.585. For a recent review of theories and
experiments on the glass transition in colloids see Sciortino and
Tartaglia\cite{952}. Mode coupling theory predicts that $D_l$ goes to zero at a
critical volume fraction $\phi_c=0.516$ following a power law:
\begin{equation}
D_l\propto(\phi_c-\phi)^{\gamma}
\label{e.1}
\end{equation}
The pictorial view is that particles become trapped in cages
formed by neighbouring particles. Experiments \cite{994} and computer
simulations \cite{907,992} confirmed this behaviour over a range of $\phi$ though with
a larger value of $\phi_c$ (0.585). However, the diffusion coefficient
is not truly zero at $\phi_c$, and some mobility is still possible
between $\phi_c$ and the volume fraction of random close packing ($\phi_{cp}=0.64$).
The reason is that fluctuations of the cage size around the average
value, allow particles to "hop" from one cage to another. Mode
coupling theory uses the static structure factor as input and
therefore cannot include the effect of fluctuations and
heterogeneity.

Introducing reversible bond formation between the
hard spheres has two consequences. Firstly, the structure factor
is modified, because on average more particles will be within each
others bond range. Secondly, the diffusion of bound particles
becomes correlated for some duration that is related to the bond
life-time. Obviously, the latter effect leads to a decrease of $D_l$,
because clusters of bound particles move more slowly than
individual spheres and not at all in the case when they form a percolating
network. However, modification of the structure may also lead to an
increase of $D_l$ compared to the hard sphere case \cite{907,926,923}, because the accessible volume for a particle in
which it can move increases. In other words, the average cage size
increases. This effect can be clearly seen for the self diffusion
of tracer particles in a medium of fixed spherical obstacles. For
a given volume fraction of obstacles, $D_l$ is smaller when the
obstacles are randomly distributed than when they form a
percolating network \cite{tracer}. 

An increase of $D_l$ with increasing attraction has actually been
observed for concentrated suspensions of hard spheres with weak
short range attraction where the particles can freely move within
the attraction range. With further increase of the interaction
energy the effect on the structure weakens, but the bond
life-time increases so that $D_l$ decreases again. These effects have
also been found in molecular dynamics simulations of such systems \cite{907,926,923}.
Mode coupling theory describes this behaviour purely on the basis
of the changes in the static structure factor and predicts
complete arrest at finite interaction strength even at low volume
fractions \cite{1029}. Systems arrested by attraction are called attractive
glasses \cite{1030} to distinguish them from the ordinary repulsive glasses \cite{994}.

Obviously, irreversible aggregation, i.e. infinite attraction
strength, leads to arrest at any volume fraction when all spheres
become part of the percolating network \cite{802}. Generally, arrested
systems at low volume fractions due to irreversible aggregation
are called gels. In analogy, percolated systems with finite
interaction strength may be called transient gels, though in
practice one would only do that only if the bond life-time is very long
so that the system flows very slowly. The issue addressed here is
how $D_l$ varies with increasing attraction strength and specifically
whether $D_l$ becomes zero in transient gels with a finite bond
life-time. In other words, whether transient gels can be
attractive glasses. Even if, as for repulsive glasses, $D_l$ does not
truly become zero due to hopping processes, the question remains
whether $D_l$ can be usefully described as going to zero  at a finite
critical interaction energy  following a power law over a broad range of $D_l$:
\begin{equation}
D_l \propto(u-u_c)^{\gamma}
\label{e.2}
\end{equation}
as was suggested in the literature 
\cite{936,923} ($u_c$). 

As mentioned above, phase separation occurs when the attraction is
strong, and needs to be avoided if one wishes to study the effect
of strong attraction on $D_l$ at low volume fractions. This can be
done to some extent by limiting the maximum number of bound
neighbours below 6 \cite{936,884} or by introducing a
correlation between the bond angles with different neighbours \cite{921}. 
One can also introduce a long range repulsive
interaction to push phase separation to lower concentrations
\cite{923,885,826}. Phase separation
disappears completely if the bonds are rigid and the bond range is
zero, because only binary collisions can occur so that the average
coordination number cannot exceed two \cite{901}. For this model
$D_l$ can be studied at any interaction strength and volume fraction
without the interference of phase separation. 

\begin{figure}
\resizebox{0.45\textwidth}{!}{\includegraphics{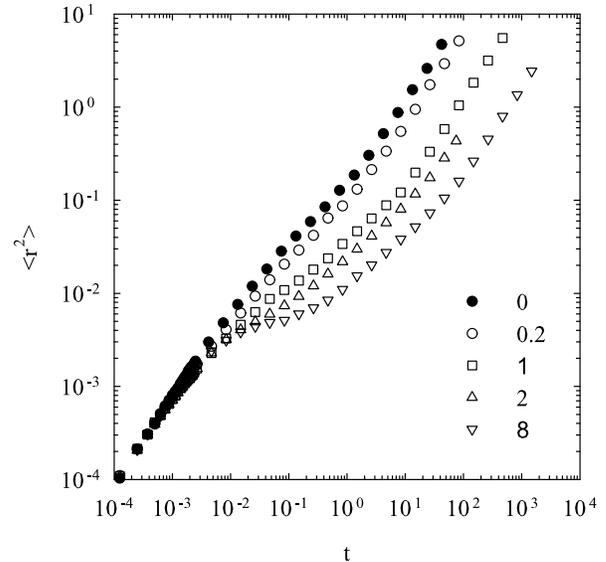}}
\caption{MSD of spheres starting from a random 
position at $\phi=0.49$ for different $t_e$ as indicated in the figure.}\label{f.1}
\end{figure}

An extensive study of the structural properties of such systems was published earlier
\cite{901}. It was shown that the equilibrium properties depend
on the ratio $\alpha/\beta $ and the elementary step size $s$ of the Brownian
motion. As long as $s$ is much smaller than the average distance
between the particles, the equilibrium properties are determined
by a single parameter called the escape time ($t_e$) that is
a combination of $s$, $\alpha$ and $\beta$. $t_e$ is defined as the excess time it
takes for two particles at contact to become decorrelated compared
to the situation where no bonds are formed ($\alpha=0$). If the
interaction range is finite and if $s$ is much smaller than the
interaction range, then the escape time becomes independent of $s$
and the equilibrium properties are the same as for spheres
interacting with a square-well potential. 

Here we present results
of the mean square displacement (MSD) of the particles as a
function of the interaction strength for systems forming rigid bonds with zero and
finite interaction strength. The main conclusion of this work is
that introducing attraction leads to a decrease of $D_l$ without any
sign of arrest at finite interaction strength. As mentioned above,
there are two different mechanisms for slowing down: one is
controlled by the bond life-time and the other is controlled by
crowding. We show here that for rigid bond formation with $\alpha=1$ the
two mechanisms are independent and can be factorized. Since the
first mechanism leads to complete arrest only for irreversible
aggregation, we propose that the expression "attractive glass" is
abandoned in favour of the expression "gel" that is commonly used
to describe irreversibly bound percolating networks. The results
presented here resemble to some extent those obtained recently by
Zaccarelli et al.\cite{936} for attractive spheres with limited
valency. The main difference is that in those simulations the
bonds were not rigid and localized motion was possible even at
infinite bond strength. We will briefly discuss the effect of bond
flexibility.

\section{Simulation method}

The simulation method for aggregation with zero interaction range
has already been detailed elsewhere \cite{901}. Initially, $N$
non-overlapping spheres with unit diameter are positioned randomly
in a box with size $L$, using periodic boundary conditions, so that
$\phi=(\pi/6)N/L^3$. Different sizes between $L=10$ and $L=50$ were used, and the
results shown here are not influenced significantly by finite size
effects. $N$ times a particle is chosen and moved a distance $s$ in a
random direction. When the movement leads to overlap with another
sphere it is truncated at contact. After this
movement step all spheres in contact are bound with probability $\alpha$
leading to the formation of $N_c$ clusters that are defined as sets
of bound particles. In the next movement step, $N_c$ times a cluster
is chosen and moved a distance $s$ in a random direction with a
probability that is inversely proportional to the diameter of the
cluster, thus simulating non-draining within the
clusters. In the following cluster construction step, all bound
particles are broken with probability $\beta$ and bonds are formed for
new contacts with probability $\alpha$. Movement and cluster formation
steps are repeated until all particles have formed a single
cluster for irreversible aggregation or until equilibrium is
reached for reversible aggregation.

\begin{figure}
\resizebox{0.45\textwidth}{!}{\includegraphics{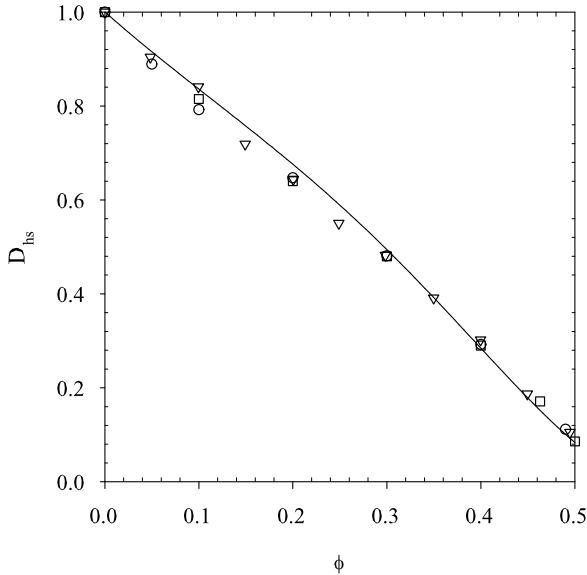}}
\caption{Dependence of the self diffusion coefficient of randomly distributed hard spheres on the volume fraction obtained from computer simulations: 
present work (circles), \cite{877} (solid line), \cite{1009} (squares) and \cite{1010} (triangles).}\label{f.2}
\end{figure}

The diffusion coefficient was calculated as $\langle{r^2}\rangle/(6t)$ and the 
time unit was defined as the time needed for a single particle to diffuse its 
own diameter so that the diffusion coefficient at infinite dilution is $D_0=1/6$. 
In this article we will present diffusion coefficients normalized by $D_0$. For the case of irreversible ($\beta=0$) diffusion 
limited ($\alpha=1$) aggregation, the kinetics of cluster growth were the 
same as predicted from the Smolechowski equations for DLCA if $s$ was chosen 
sufficiently small\cite{802}. As mentioned in the introduction, the equilibrium 
properties for reversible aggregation with zero interaction range are determined by the escape time \cite{901}:
\begin{equation}
t_e=(2.37s+s^2)\frac{\alpha}{\beta}
\label{e.3}
\end{equation}  
At equilibrium, a distribution of self-similar transient clusters
is formed together with a transient percolating network if $t_e$ 
and $\phi$ are sufficiently large.

\begin{figure}
\resizebox{0.45\textwidth}{!}{\includegraphics{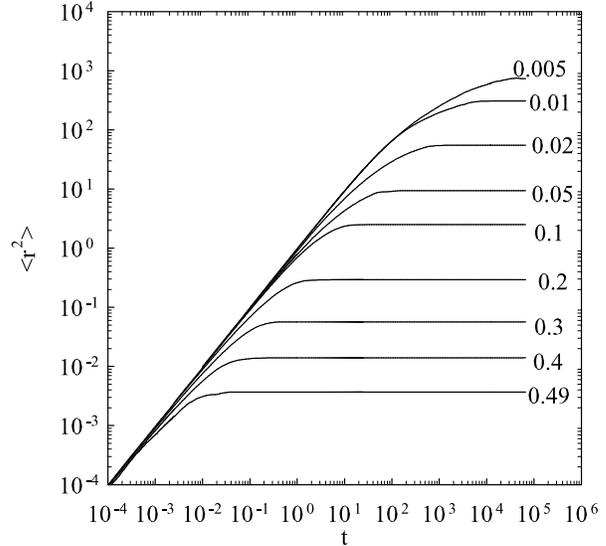}}
\caption{MSD of spheres during irreversible 
aggregation (DLCA) for different volume fractions as 
indicated in the figure.}\label{f.3}
\end{figure}

Reversible aggregation with finite interaction range $\varepsilon$ was 
simulated by forming bonds with probability $\alpha$ when the centre 
to centre distance between particles is less than $1+\varepsilon$ 
\cite{955}. Bonds were again broken with probability $\beta$. In 
this case the equilibrium properties are independent of the 
step size if $s\ll\varepsilon$. The system is equivalent to particles 
interacting through a square well attraction with interaction 
energy $u=\ln(1-P)$, where $P$ is the probability that particles within 
each others interaction range are bound: $P= \alpha/(\alpha +\beta)$. 
$u$ is given in units of the thermal energy and is equivalent to the 
inverse temperature that is sometimes used to express the 
interaction strength. The equilibrium properties of systems with 
different interaction range are close if they are compared at the 
same second virial coefficient ($B_2$)\cite{787}. $B_2$ is the sum of the 
excluded volume repulsion and the square well attraction: 
$B_2=B_{rep}-B_{att}$. $B_{rep}=4$ and $B_{att}$ is determined both 
by the interaction strength  and the interaction range  \cite{892}:
\begin{equation}
B_{att}=4[(1+\varepsilon)^3-1][\exp{(-u)}-1]
\label{e.4}
\end{equation}  
in units of the particle volume

\section{Results}

\subsection{Zero interaction range} 
Fig. \ref{f.1} shows the mean square
displacement, i.e. $\langle{r^2}\rangle$, of spheres at $\phi=0.49$ 
for different values of $t_e$ with $\alpha=1$, starting from 
a random distribution. Initially, the particles move freely 
until they begin colliding with other particles. This leads 
to slowing down of the MSD even in the
absence of attraction ($t_e=0$). At long times $\langle{r^2}\rangle$ is again
proportional to $t$, but the long time diffusion coefficient ($D_l$) is
smaller. Fig. \ref{f.2} shows that the dependence of the normalized
diffusion coefficient of hard spheres in the absence of attraction
($D_{hs}$) on $\phi$ is in good agreement with literature results \cite{877,872,1009,1010,993}.

\begin{figure}
\resizebox{0.45\textwidth}{!}{\includegraphics{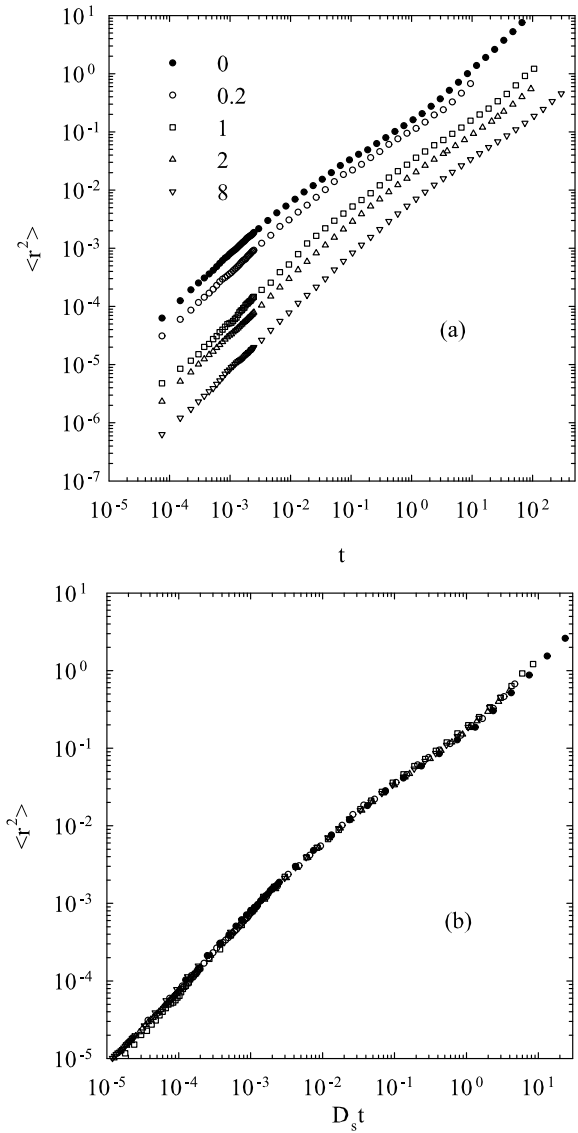}}
\caption{MSD for equillibriated systems ($a$) at $\phi=0.49$ for different 
$t_e$. Fig. 4$b$ shows a mastercurve obtained by plotting the same data as a function of $D_st$.} \label{f.4}
\end{figure}

Introducing attraction slows down the MSD, because transient
clusters are formed that move more slowly and above a critical value
of $t_e$
 a transient percolating network is formed. Particles that are part of the transient network are
immobile and need to break bonds before they can diffuse. In the
limit of $t_e=\infty$, corresponding to irreversible DLCA, all particles are
permanently stuck when they become part of the percolating
network. Naturally, the value of $\langle{r^2}\rangle$ where the particles get stuck
decreases with increasing volume fraction, see Fig \ref{f.3}. 

Once the
system has reached equilibrium one can start again measuring the
MSD. Fig. \ref{f.4}$a$ shows the MSD for the same systems as in Fig \ref{f.1},
but starting from the equilibrium state. In this case the 
initial diffusion is slower than that of individual spheres, since
the system contains clusters and for $t_e>0.43$  a percolating network.
After some time the clusters collide causing a further slowing
down of the MSD, but simultaneously, for $\alpha=1$, bonds are formed and
broken. Of course, for $t\gg t_e$ the same long time diffusion
coefficient is obtained independent of the starting configuration,
compare Fig. \ref{f.1}. Remarkably, the effect of particle collisions on
the slowing down of the MSD is the same at different $t_e$. Therefore
the MSD obtained at different $t_e$ can be superimposed within the
statistical error by simple time shifts (see Fig. \ref{f.4}$b$). The
implication is that $D_l$ can be factorized in terms of the short time
diffusion coefficient ($D_s$) and $D_{hs}$:
\begin{equation}
D_l(t_e,\phi)=D_s(t_e,\phi) \cdot D_{hs}(\phi)
\label{e.5}
\end{equation}  

At short times, i.e. before bond breaking becomes
significant and before the particles collide, the MSD is determined by
free diffusion of the clusters. Consequently, $D_s$ can be calculated from the
size distribution of the clusters ($N(m)$) which is a function of
$t_e$ and $\phi$:
\begin{equation}
D_s(t_e,\phi)=\sum_{m}\frac{m\cdot N(m)\cdot D(m)}{N}
\label{e.6}
\end{equation}  
where $D(m)$ is the average free diffusion coefficient of clusters with
aggregation number $m$. If a percolating network is present we need
to consider only the sol fraction of mobile clusters ($F$):
 \begin{equation}
D_s=\frac{F}{m_n} \cdot \frac{\sum_{m}^{sol}m\cdot N(m)\cdot D(m)}{\sum^{sol}_{m}N(m)}
\label{e.7}
\end{equation} 
where $m_n$ is the number average aggregation number of the sol fraction. In the
absence of hydrodynamic interactions $D(m)\propto m^{-1}$ so that $D_s\propto m_n^{-1}$. 
For the more realistic case of non-draining clusters $D(m)$ is
inversely proportional to the radius of the clusters so that $D_s$
decreases more weakly with increasing $m_n$. 

\begin{figure}
\resizebox{0.45\textwidth}{!}{\includegraphics{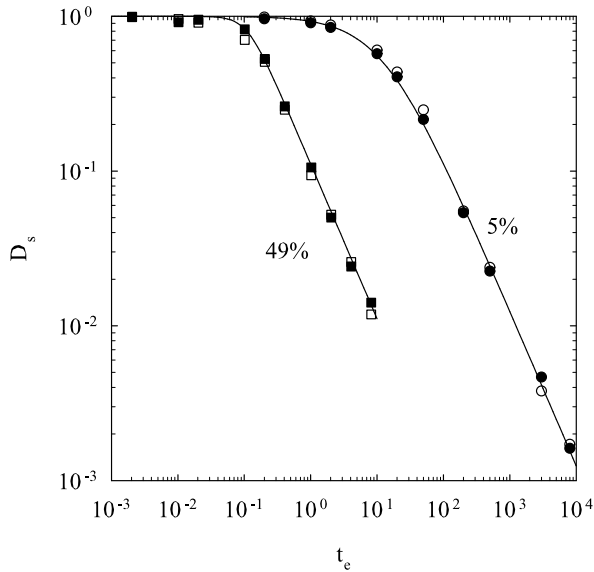}}
\caption{The short time diffusion coefficient as a 
fuction of $t_e$ 
for $\phi=0.49$ (squares) and $\phi=0.05$ (circles) . Closed and open symbols represent simulation
results and calculations using Eq. (\ref{e.6}).}\label{f.5}
\end{figure}

Figure \ref{f.5} shows $D_s$ as a
function of $t_e$ for  $\phi=0.49$ and  $\phi=0.05$. We have calculated $D_s$ from
the cluster size distribution using Eq. (\ref{e.6}) and $D_l$ was calculated
using Eq. (\ref{e.5}). Comparison with the simulation results, see Fig (\ref{f.5}),
demonstrates that the effect of attraction on $D_s$ is indeed fully
determined by the cluster size distribution. Initially, $D_l$
decreases with increasing $t_e$, because $m_n$ increases until at a
critical value of $t_e$ ($t_e^{*}$) a percolating network is formed of
immobile particles. $t_e^{*}$ is $0.43$ and $180$ for $\phi=0.49$ and $\phi=0.05$, repectively. At the percolation threshold the weight
average aggregation number diverges \cite{901}, but $m_n$ and thus $D_l$
remain finite. The maximum value of $m_n$ is obtained at the
percolation threshold, beyond which it decreases with increasing volume
fraction. Beyond the percolation threshold the sol fraction
decreases much more strongly with increasing attraction than $m_n$ so
that $D_l$ continues to decrease. However, it is obvious that $D_l$ will
become zero only if $F=0$, i.e. at infinite attraction strength. At large values of $t_e$, $D_l$ decreases linearly with increasing
$t_e$. As expected, at lower volume
fractions stronger attraction is needed to cause significant
slowing down.

\begin{figure}
\resizebox{0.45\textwidth}{!}{\includegraphics{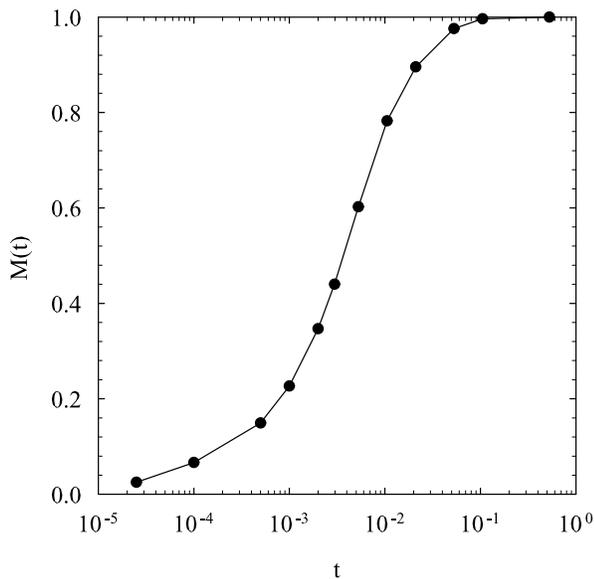}}
\caption{Fraction of mobile particles as a fuction of time for an equillibrated system 
at $t_e=2$ and $\phi=0.49$ . The solid line 
is a guide to the eye.}\label{f.6}
\end{figure}

The MSD shown in Fig. \ref{f.4} represent an average over all particles, 
and do not show how the displacement of the particles is distributed.
In the absence of attraction the probability distribution 
that a particle has moved a distance $r^2$ at a given time ($P(r^2)$) is 
given by:
\begin{equation}
P(r^2,t)=\frac{2\pi}{(4\pi Dt)^{3/2}}r\exp(-\frac{r^2}{4Dt})
\label{e.8}
\end{equation}
because each particle is equivalent. However, in the presence 
of attraction the displacement is highly heterogeneous, because the 
particles belong to clusters of different sizes. We have illustrated 
this for $\phi=0.49$ at $t_e=2$. In this case almost all of the particles 
belong to the gel fraction and cannot move until their bonds 
are broken. Thus at short times the displacement is highly 
heterogeneous with a large fraction of particles that do not move at 
all and a small fraction of sol particles ($F$) that diffuse freely 
until they collide. With increasing time, more and more 
of the sol particles collide with the percolating network 
and stop moving, while more and more gel particles break 
lose and start diffusing. Fig \ref{f.6} shows how the 
fraction of mobile particles that has moved after a time 
$t$ ($M(t)$) increases with increasing $t$.

\begin{figure}
\resizebox{0.45\textwidth}{!}{\includegraphics{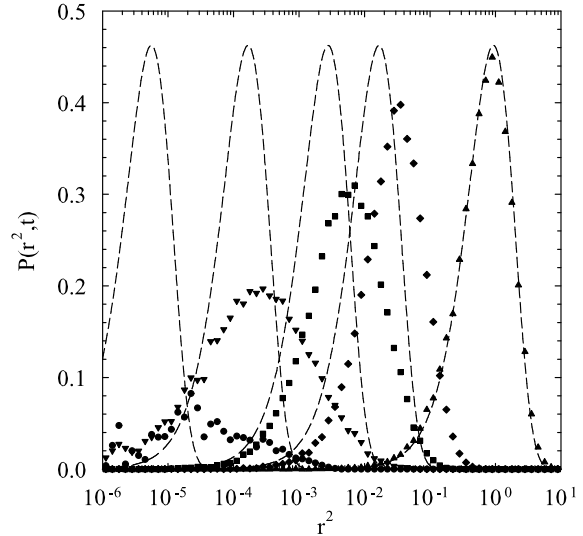}}
\caption{Probability distribution of $r^2$ at different times 
$0.001$ (circle), $0.3$ (triangledown), $0.5$ (square), $3$ (diamond) and $167$ 
(triangle) for $\phi=0.49$ and $t_e=2$. 
The solid lines represent the distributions for freely diffusing spheres 
with the same average MSD, see Eq. \ref{e.8}.}\label{f.7}
\end{figure}

Figure \ref{f.7} compares $P(r^2)$ at different times with 
the distribution that would have been found for same $\langle r^2\rangle$ if all particles 
had been equivalent. Note that the area under the curves is equal to 
$M(t)$. The heterogeneity of the displacement decreases with 
increasing time until for $t\gg t_e$ all particles have 
formed and broken bonds many times so that $M(t)=1$ and 
$P(r^2)$ is given by Eq. \ref{e.8} . Similar observations were made at $\phi=0.05$.

Figure \ref{f.7} resembles the results presented by
Puertas et al. \cite{925}. However, they used molecular dynamics simulations
for which bound particles move freely as long as the displacement
does not involve breaking of bonds, as will be discussed below.
For this reason they observed a peak situated at a small $r$ that
represents the displacement of bound particles. A second peak was found at larger values of $r$ representing
the sol fraction and moved to longer distances with increasing
time. The amplitude of this peak increased with increasing time as
more and more particles break their bonds. 

\begin{figure}
\resizebox{0.45\textwidth}{!}{\includegraphics{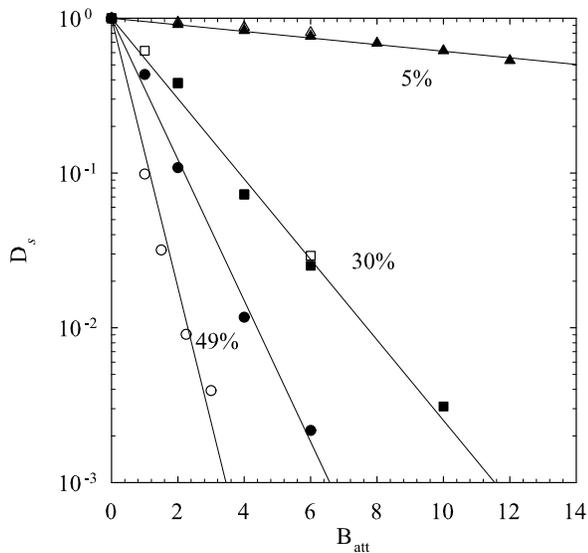}}
\caption{The short time diffusion coefficient at 
$\phi=0.49$ (circles), $\phi=0.3$ (squares) and $\phi=0.05$ 
(triangles) for two different interaction ranges $\varepsilon=0.5$ (closed) 
and $\varepsilon=0.1$ (open) as a function of $B_{att}$. The solid lines are guide to the eye.}\label{f.8}
\end{figure}

\subsection{Finite interaction range}

The effect of finite interaction range was tested for $\varepsilon=0.1$
and $\varepsilon=0.5$. We found also for this case that $D_l$ could be factorized
into $D_s$ and $D_{hs}$, see Eq. \ref{e.5}, and that the effect of attraction on $D_l$
is fully determined by the free diffusion of the clusters.
However, the dynamic heterogeneity that was important for the zero
range interaction disappeared rapidly with increasing interaction
range especially at higher concentrations, because monomers and
clusters can move only a short distance before they interact.
Individual particles exchange rapidly between different clusters
including the gel fraction, at least for $\alpha=1$. Therefore each
particle explores more rapidly the different dynamics of the
system. 

Recently, it was shown that the cluster distribution in
equilibrium is similar for the two interaction ranges if compared
at the same values of the second virial coefficient, at least for 
$\phi$ up to $0.2$ \cite{955}. We therefore expect that $D_s$ is the same at the same $B_2$.
Fig. \ref{f.8} shows $D_s$ as a function of $B_{att}$ for 3 different volume
fractions with $\varepsilon=0.5$ and $\varepsilon=0.1$. As 
mentioned in the introduction, for lower volume fractions 
the slowing down of $D_s$ can be studied only
over a limited range before phase separation occurs. As
expected, $D_s$ decreases with increasing $B_{att}$ and the decrease is more
important at higher concentrations. Again no sign of critical
slowing down of the MSD was observed. For $\varepsilon=0.5$ and $\varepsilon=0.1$ the dependence of $D_s$ on $B_{att}$ was
similar  for $\phi=0.05$ 
and $0.3$, but for $\phi=0.49$ it
was stronger for $\varepsilon=0.1$ than for $\varepsilon=0.5$. At high volume
fraction $B_2$ is no longer the main parameter that determines the
cluster size distribution because higher order interaction becomes
important. 

\begin{figure}
\resizebox{0.45\textwidth}{!}{\includegraphics{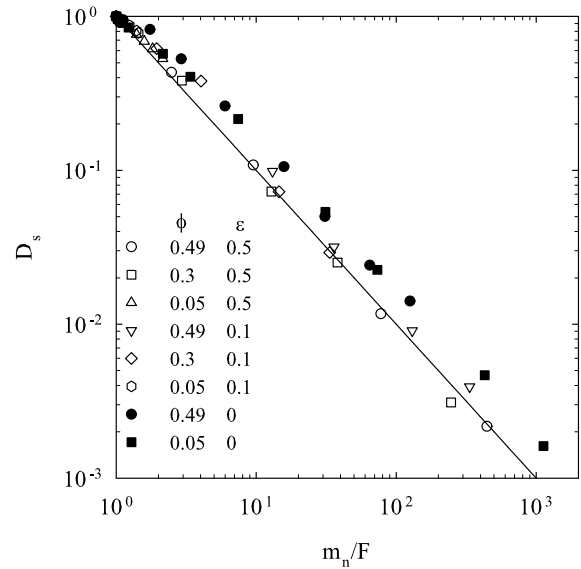}}
\caption{The short diffusion coefficient as a function $m_n/F$
for different volume fractions ($\phi$) and interaction ranges ($\varepsilon$) as 
indicated in the figure . The solid 
line represents $D_s=m_n/F$ }\label{f.9}
\end{figure}

Recently, Foffi et al. \cite{912} reported a simulation study of
the effect of the interaction range on $D_l$ for hard spheres
with a square well interaction. Molecular dynamics simulations
were used that gave the same equilibrium structures as with the method
used here. However, molecular dynamics simulations gives different dynamics so that absolute values of $D_l$ cannot
be compared with the results presented here. Nevertheless, they also found that the
MSD of particles was independent of the interaction range if
compared at the same value of $B_2$ and they argued that $D_l$ was the same
because the number of bonds and the bond life-time was determined
by $B_2$. We expect that deviations will be found also with this
method at higher concentrations and larger interaction ranges when
higher order interaction becomes important.

\section{Discussion}

There are two causes for the decrease of $D_l$ in a system of
attractive hard spheres. The first one is collisions with other
spheres. This effect of crowding increases with increasing volume
fraction and leads to strong decrease of $D_l$ near $\phi \approx 0.58$ that can be
described to some extent by mode coupling theory, as mentioned in
the Introduction. The second cause is bond formation, which leads
to the formation of transient clusters and gels. This effect
increases with increasing attraction. The diffusion at short times
is not influenced by collisions so that in the absence of
attraction $D_s$ is equal to the free diffusion of the particles. The
decrease of $D_s$ with increasing attraction is caused solely by
cluster formation and can be calculated from the cluster size
distribution. Since for any finite interaction there is a finite
fraction of free particles and clusters, $D_s$ only becomes zero when
the interaction is infinitely strong. The decrease of $D_s$ with increasing interaction strength 
is mainly determined by $m_n/F$, see Fig. \ref{f.9} and it is dominated by the decrease of the sol fraction for strong attraction when $m_n$ is close to unity. 

The subsequent decrease of the diffusion coefficient from $D_s$ to $D_l$ at long
times is caused by crowding. An important observation is that the
slowing down caused by crowding is independent of the attraction
strength for $\alpha=1$. The reason is that by definition for reversible
DLCA the reversibility is expressed as soon as collisions occur
and the memory of the connectivity is lost. Consequently,
diffusion is slowed down by attraction, but does not become
zero as long as the attraction is finite. The situation is
different for reversible RLCA in which case the collisions
occur between long-lived clusters and between the clusters and the percolating
network. For given attraction strength $D_s$ is the same for reversible DLCA and RLCA, but for RLCA
the ratio $D_l/D_s$ decreases with decreasing bond formation probability. This
situation will be explored elsewhere. 

\begin{figure}
\resizebox{0.45\textwidth}{!}{\includegraphics{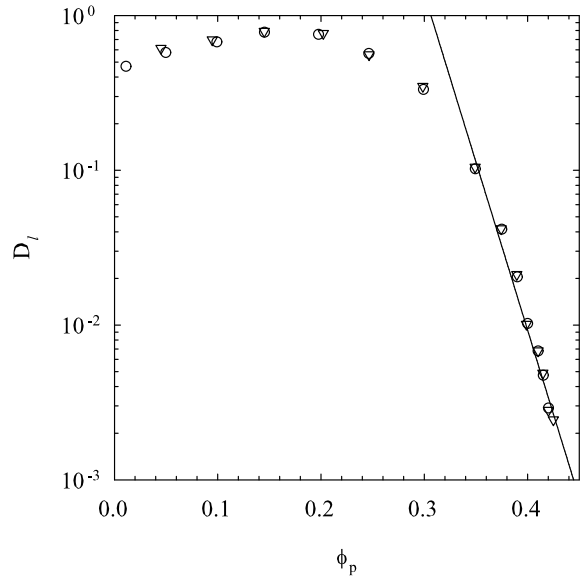}}
\caption{Long time self diffusion coefficient as a fuction of 
($\phi_{p}$) for 2 different values of $\phi_{p}^{G}$ 
(0.4265 (circles) and 0.4519 (triangles)) from \cite{923}. The solid lines represents 
$D_l \propto \exp(-50\phi_p).$}\label{f.10}
\end{figure}

The results presented here apparently
contradict molecular dynamics simulations that showed critical
slowing down at finite interaction strength \cite{883,923}. 
The main difference between our simulation method and
molecular dynamics is that rigid bonds are formed so that only
cluster motion is possible. In the molecular dynamics simulations
bound particles are still allowed to diffuse freely as long as no
bonds are broken. The implication is that $D_s$ is equal to the free particle diffusion independent of the
interaction strength. Consequently $D_l$ is
faster and $D_l$ is even larger than $D_{hs}$ for weak attraction. We have
included bond flexibility in the simulations with finite
interaction strength by allowing free diffusion for bound
particles as long as it does not lead to bond breaking. Details of
these simulations will be reported elsewhere. Here, we only
mention that including bond flexibility increases $D_l$. It is clear, that allowing more
freedom for movement does not lead to a critical slowing down of
the diffusion at a finite interaction range so that the apparent contradiction with molecular dynamics simulations persists. 
But, a close look at the molecular
dynamics simulation results shows that
they can also be interpreted in terms of a power law decrease. For
instance, Puertas et al. \cite{923} simulated hard spheres with a short range
depletion interaction caused by the addition of polymers. In
Fig. \ref{f.10} we have replotted $D_l$ as a function of the polymer volume
fraction ($\phi_p\propto -u$) for
two different particle volume fractions. The authors interpreted
the data in terms of Eq.\ref{e.2}. At each concentration the smallest values of $D_l$ deviated from
this expression, which was attributed to "hopping". It is clear,
however, from Fig. \ref{f.9} that the decrease of $\ln D_l$ at large $\phi_p$ can also be
described as a power law even for the smallest value of $D_l$ implying that
$D_l=0$ only at infinite attraction. 

Very recently, Zaccarelli et al. \cite{936} 
made a detailed study of the dynamics in attractive hard
sphere systems with limited valence (3 and 4) using molecular
dynamics simulations. They observed that for strong attraction 
the variation of $\ln D_l$ could be described by a power law in
terms of $u$ at least for $\phi\leq 0.55$ implying that arrest only 
occurred at infinite attraction. On the other hand they could
describe $D_l$ as a function of $\phi$ in terms of Eq. \ref{e.1}. $\phi_c$ was almost
constant when increasing the attraction and only for strong
attraction did they find a weak decrease of $\phi_c$, but the<
extrapolation was uncertain in this case. The authors used the
expression reversible gel for systems that arrested at $u\rightarrow -\infty$ $(T\rightarrow 0)$
and glass (attractive or repulsive) for systems that arrested at 
$\phi\rightarrow \phi_c$. The expression attractive glass was introduced to describe the
arrest that occurs for a given volume fraction at a finite
attraction energy. It is clear from the present study that attractive
glasses in this sense are only formed for irreversible
aggregation.

\section{Conclusion}

Reversible cluster aggregation of hard spheres leads to
equilibrium systems containing transient clusters and, above a
critical interaction strength, a transient percolating network. If
the aggregation is diffusion limited and the bonds are rigid, then
the effect of attraction on $D_l$ is decoupled from the effect of
crowding. The latter is equal to that of non-interacting hard
spheres, while the former is fully determined by the cluster size
distribution. The self diffusion coefficient of the spheres
decreases with increasing attraction, but becomes zero only for
irreversible aggregation, contrary to predictions from mode
coupling theory. Therefore attractive glasses in the sense of
systems that are dynamically arrested by a finite interaction
energy do not exist.

\begin{acknowledgments}
This work has been supported in part by a grant from the Marie Curie Program
of the European Union numbered MRTN-CT-2003-504712.
\end{acknowledgments}

\end{document}